\documentclass[12pt,fleqn]{article}
\usepackage{graphicx}
\usepackage{latexsym}
\usepackage{amsfonts}
\usepackage{amssymb}
\usepackage{amsmath}

\newcommand{\bea}{\begin{eqnarray}}
\newcommand{\eea}{\end{eqnarray}}
\newcommand{\bef}{\begin{figure}}
\newcommand{\enf}{\end{figure}}
\newcommand{\ball}{\begin{array}{ll}}
\newcommand{\bal}{\begin{array}{l}}
\newcommand{\ea}{\end{array}}
\newcommand{\feta}{{\boldsymbol{\eta}}}
\newcommand{\falpha}{{\boldsymbol{\alpha}}}

\newcommand{\mubo}{{\boldsymbol{\mu}}}
\newcommand{\rhobo}{{\boldsymbol{\rho}}}
\newcommand{\grad}{{\boldsymbol{\nabla}}}
\newcommand{\rel}{{\mathbb{R}}}
\newcommand{\nat}{{\mathbb{N}}}
\newcommand{\ganz}{{\mathbb{Z}}}
\newcommand{\ord}{{\mathcal{O}}}

\newcommand{\abo}{{\mathbf{a}}}
\newcommand{\bbo}{{\mathbf{b}}}
\newcommand{\xbo}{{\mathbf{x}}}
\newcommand{\ubo}{{\mathbf{u}}}
\newcommand{\Fbo}{{\mathbf{F}}}

\newcommand{\ybo}{{\mathbf{y}}}
\newcommand{\zbo}{{\mathbf{z}}}
\newcommand{\mbo}{{\mathbf{m}}}
\newcommand{\kbo}{{\mathbf{k}}}
\newcommand{\ebo}{{\mathbf{e}}}
\newcommand{\Rbo}{{\mathbf{R}}}
\newcommand{\Jbo}{{\mathbf{J}}}
\newcommand{\Mbo}{{\mathbf{M}}}

\newcommand{\Di}{{\stackrel{\circ}{D}}}

\numberwithin{equation}{section}

\title{Stationary measures and hydrodynamics of zero range processes with several species of particles}
\author{Stefan Gro\ss kinsky and Herbert Spohn\footnote{Zentrum Mathematik, Technische Universit\"at M\"unchen, 85747 Garching bei M\"unchen, Germany; e-mail: stefang@ma.tum.de, spohn@ma.tum.de}}
\date{\today}

\begin{document}
\maketitle

\begin{abstract}
We study general zero range processes with different types of particles on a $d$-dimensional lattice with periodic boundary conditions. A necessary and sufficient condition on the jump rates for the existence of stationary product measures is established. For translation invariant jump rates we prove the hydrodynamic limit on the Euler scale using Yau's relative entropy method. The limit equation is a system of conservation laws, which are hyperbolic and have a globally convex entropy. We analyze this system in terms of entropy variables. In addition we obtain stationary density profiles for open boundaries.
\end{abstract}

\section{Introduction}
The zero range process is a stochastic particle system on the $d$-dimensional lattice $\mathbb{Z}^d$ where the jump rate $g(k)$ of a given particle depends only on the occupation number $k$ at its current position. This model was originally introduced as a simple example of an interacting Markov process \cite{spitzer70}. Various properties have been established, among them the existence of the dynamics under very general conditions, classification of invariant measures, and hydrodynamic limits \cite{andjel82,kipnislandim,liggettetal81}.

In this paper we generalize the zero range process to $n$ different types of particles. The jump rate $g_i$ of the $i$-th component depends on the occupation numbers of all $n$ species at a given site. For the process to have stationary product measures, these rates cannot be chosen arbitrarily. We find that such measures exist if and only if the logarithm of the jump rates $g_i$ is given as the lattice derivative $\nabla_i G$ of a suitable potential $G$. In general, the stationary product measures do not factorize with respect to the different components.

Our goal is the derivation of the hydrodynamical equations on the Euler scale, which are expected to be given by a system of $n$ conservation laws. For this purpose we use the relative entropy method of Yau \cite{yau91}, which, while directly applicable to the present case, has the disadvantage to yield the desired result only up to the first shock. So far we did not attempt to extend our result to all times following e.g.\ the lines in \cite{rezakhanlou91}. We show that the system of conservation laws is hyperbolic and the thermodynamic entropy of the stationary measure is a Lax entropy \cite{smoller}. This property follows from certain reciprocity relations for the steady currents of the components. Such relations have recently been established for one-dimensional interacting particle systems restricted to nearest neighbor interaction and finite occupation numbers \cite{tothetal02}.

One source of motivation for this work comes from the analysis of one-dimensional driven diffusive systems with open boundaries. For one-component systems the analysis of the hydrodynamic limit equation led to the theory of boundary-induced phase transitions which provides a general framework for a quantitative description of the steady-state selection in systems which are in contact with particle reservoirs at their boundary \cite{krug91,popkovetal99}. In systems with more than one conserved quantity interesting new phenomena have been found such as phase separation and spontaneous symmetry breaking \cite{evansetal95,evansetal98}, for a recent review see \cite{schuetz03}. Again it is natural to ask for principles governing steady-state selection and the resulting phase diagram in systems with many species of particles. The macroscopic behavior of such systems has been examined to some extent \cite{popkovetal02} only recently and there are very few rigorous results \cite{tothetal02,tothetal02b}. In our note we analyze the system of conservation laws for open boundaries in terms of entropy variables \cite{ulbrich}. For stationary solutions the system decouples, and we are able to derive stationary density profiles, which we state explicitly for one-dimensional geometry.

\section{Stationary product measures}
  \subsection{The $n$-component zero range process}
Let us consider a zero range process with translation invariant jump rates on the $d$-dimensional torus $\Lambda_L =(\ganz /L\ganz )^d$. There are $n$ different species of particles and let $\eta_i (\xbo ) \in\nat$ be the number of particles of component $i\in\{ 1,\ldots ,n\}$ on site $\xbo\in\Lambda_L$, where $\nat =\{ 0,1,2,\ldots \}$. The state space is given by $\Omega_L =(\nat^n )^{\Lambda_L}$ and we denote a particle configuration by $\feta =\big(\feta (\xbo )\big)_{\xbo\in\Lambda_L} =\Big(\big(\eta_1 (\xbo ),\ldots ,\eta_n (\xbo )\big)\Big)_{\xbo\in\Lambda_L}$. At a given site $\xbo\in\Lambda_L$, the number $\eta_i (\xbo )$ of $i$-type particles decreases by one after an exponential waiting time with rate $g_i \big(\feta (\xbo )\big)$ and the leaving particle jumps to site $\xbo+\ybo$ with probability $p_i (\ybo )$. The jump probabilities $p_i :\ganz\to [0,1]$ are normalized and assumed to be of finite range $R\in\ganz^+$, i.e.
\bea\label{passum}
\sum_\ybo p_i (\ybo )=1\ ,\quad p_i (\mathbf{0})=0\quad\mbox{and}\quad p_i (\ybo )=0\mbox{ for }|\ybo |>R\ .
\eea
To exclude hidden conservation laws the $p_i$'s have to be irreducible, so that every particle can reach any site of the lattice with positive probability. The rate function $g_i :\nat^n \to [0,\infty )$ vanishes for all $\kbo =(k_1 ,\ldots ,k_n )\in\nat^n$ with $k_i =0$ and is otherwise positive and uniformly bounded from below, i.e. using the shorthand $\kbo =(\kbo^i ,k_i )$,
\bea\label{gassum}
g_i (\kbo^i ,k_i )=0\,\Leftrightarrow\, k_i=0\quad\mbox{and}\quad g_i^* :=\sup_{n\in\nat}\inf_{|\kbo |\geq n\atop k_i >0} g_i (\kbo^i ,k_i )>0
\eea
for all $i=1,\ldots ,n$. With these assumptions the generator of the zero range process is given by
\bea\label{generator}
(\mathcal{L} f)(\feta )=\sum_{\xbo ,\ybo\in\Lambda_L } \sum_{i=1}^n g_i \big(\feta (\xbo )\big)\, p_i (\ybo )\Big( f(\feta^{i;\xbo ,\xbo +\ybo} ) -f(\feta )\Big) ,
\eea
regarded as a linear operator on $C(\Omega_L ,\rel )$. The configuration $\feta^{i;\xbo ,\xbo +\ybo}$ results from $\feta$ after one particle of component $i$ is moved from $\xbo$ to $\xbo+\ybo$, i.e.\ $\eta_j^{i;\xbo ,\xbo +\ybo} (\mathbf{z}) =\eta_j (\mathbf{z}) +\delta_{ij}(\delta_{\mathbf{z},(\xbo +\ybo )mod\, L^d} -\delta_{\mathbf{z},\xbo })$ for all $\zbo\in\Lambda_L$, $j=1,\ldots ,n$. The number of particles $N_i (\feta ) =\sum_{\xbo\in\Lambda_L}\eta_i (\xbo )$ of component $i$ is conserved, and these are the only conserved quantities. They divide the configuration space into finite, invariant subsets with fixed $N_i \in\ganz^+$, $i=1,\ldots ,n$. Restricted to such a subset, $\mathcal{L}$ is a finite dimensional matrix and the process is well defined. However, for $L=\infty$ this is true only for ``reasonable'' initial conditions and under additional assumptions on the jump rates, which are given in Section 3.1 (cf.\ \cite{andjel82,liggettetal81}).

  \subsection{Existence of stationary product measures}
For the one-component process $n=1$, as well known, there exists a family of translation invariant stationary product measures (see e.g.\ \cite{andjel82,kipnislandim,evans00})
\bea\label{grandcanens}
\bar{\nu}_\mu^L (\feta )=\prod_{\xbo\in\Lambda_L}{1\over Z(\mu )}\, e^{\mu\,\eta_1 (\xbo )}\prod_{k=1}^{\eta_1 (\xbo )} {1\over g_1 (k)}\quad\mbox{with }Z(\mu )=\sum_{l=0}^\infty e^{\mu\, l}\prod_{k=1}^l {1\over g_1 (k)}\ .
\eea
The parameter $\mu\in\rel$ is the chemical potential and controls the average particle density, the normalizing constant $Z(\mu )$ is the partition function. Often $\mu$ is replaced by the fugacity $\phi =\exp [\mu ]\in [0,\infty )$. We use this notation in Section 4.2, but for Sections 2 and 3 the chemical potential turns out to be more convenient. In the case $n>1$, the stationary measures are of product form only under the following condition on the jump rates:\\
\\
\textbf{Assumption:}\it\ For every $i,j\in\{1,\ldots ,n\}$ and $\kbo =(k_1 ,\ldots ,k_n ) \in\nat^n$ with $k_i ,k_j >0$ let
\bea\label{a1}
g_i (\kbo )\, g_j (\kbo^i ,k_i -1 )=g_j (\kbo )\, g_i (\kbo^j ,k_j -1 )\ .
\eea\rm\\[-5mm]

This assumption is equivalent to the existence of a potential $G:\nat^n \to\rel$ for the logarithm of the jump rates such that
\bea\label{a2}
\log g_i (\kbo )=G(\kbo )-G(\kbo^i ,k_i -1)\ .
\eea
Given $G$, the jump rates defined via (\ref{a2}) clearly satisfy (\ref{a1}) by construction. On the other hand for given jump rates $g_i$ obeying (\ref{a1}) one can define $G$ recursively via (\ref{a2}) by fixing $G(0,\ldots ,0)=0$. This construction does not depend on the order of summation, since by (\ref{a1}) the sum over every closed path vanishes. For example one can choose
\bea\label{hprop}
G(\kbo )&=&\sum_{j_1 =1}^{k_1} \log g_1 (j_1 ,0,\ldots ,0) +\nonumber\\
& &\sum_{j_2 =1}^{k_2} \log g_2 (k_1 ,j_2 ,0,\ldots ,0)+\ldots +\sum_{j_n =1}^{k_n} \log g_n (k_1 ,\ldots ,k_{n-1},j_n)\ .
\eea
\\
\textbf{Theorem 1:} (Stationary product measures)\it\\
The zero range process defined in (\ref{generator}) with more than one component, $n>1$, has stationary product measures if and only if the condition (\ref{a1}), equivalently (\ref{a2}), is fulfilled. In this case the family of stationary measures can be written as
\bea\label{statmeas}
\bar{\nu}^L_\mubo (\feta )=\prod_{\xbo\in\Lambda_L}{1\over Z(\mubo )}\,\exp\Big[-G\big(\feta (\xbo )\big) +\sum_{i=1}^n \mu_i \,\eta_i (\xbo ) \Big]
\eea
with the chemical potentials $\mubo =(\mu_1 ,\ldots ,\mu_n ) \in D_\mu$ as parameters. $D_\mu$ is the domain of convergence of the partition function
\bea\label{part}
Z(\mubo )=\sum_{\kbo \in\nat^n} \exp\Big[-G(\kbo ) +\sum_{i=1}^n \mu_i \,k_i \Big]\ .
\eea
$D_\mu$ is a nonempty, convex subset of $\rel^n$ with infinite volume measure.\rm\\
\\
\textbf{Proof:} First we assume that (\ref{a1}) and (\ref{a2}) are satisfied and show that $\bar{\nu}^L_\mubo$ defined in (\ref{statmeas}) is stationary. Since $G$ exists by assumption, the measure $\bar{\nu}^L_\mubo$ is well defined on $D_\mu$ and $\big\{\mubo\in\rel^n \big|\mu_i <\log g_i^* ,\, i=1,\ldots ,n\big\}\subset D_\mu$ because of (\ref{gassum}). For $Z(\mubo^1 )$, $Z(\mubo^2 )<\infty$ it is easy to see that $Z(q\mubo^1 +(1-q)\mubo^2 )<\infty$ for all $q\in [0,1]$, thus $D_\mu$ is convex. The remainder of the first part of the proof is a straightforward generalization of the standard argument for $n=1$, given e.g.\ in \cite{kipnislandim}. To prove stationarity of $\bar{\nu}^L_\mubo$ we have to show that for all $f\in C(\Omega_L ,\rel )$
\bea\label{toshow}
\langle\mathcal{L} f\rangle_{\bar{\nu}^L_\mubo } &=&\sum_{\feta\in\Omega_L}\sum_{\xbo ,\ybo\in\Lambda_L}\sum_{i=1}^N g_i \big(\feta (\xbo ) \big)\, p_i (\ybo )\Big( f(\feta^{i;\xbo ,\xbo +\ybo} ) -f(\feta )\Big)\bar{\nu}^L_\mubo (\feta )=0\ .
\eea
With (\ref{a2}) it is easy to show that the one point-marginal $\bar{\nu}_\mubo$ satisfies
\bea\label{cancel}
\bar{\nu}_\mubo (\kbo )={\bar{\nu}_\mubo (\kbo^i ,k_i -1)\over g_i (\kbo )}\, \exp [\mu_i ]
\eea
for all $i=1,\ldots ,n$, $\kbo =(k_1 ,\ldots ,k_n )\in\nat^n$ with $k_i >0$. For every $\xbo ,\ybo\in\Lambda_L$ one has
\bea
\sum\limits_{\feta\in\Omega_L}\!\! g_i (\feta (\xbo ))f(\feta^{i;\xbo ,\xbo +\ybo} )\bar{\nu}^L_\mubo (\feta )=\sum\limits_{\feta\in\Omega_L}\!\! g_i \Big((\eta_1 ,..,\eta_i \! +\! 1,..,\eta_n )(\xbo\! -\!\ybo )\Big)f(\feta )\bar{\nu}^L_\mubo (\feta^{i;\xbo +\ybo ,\xbo} )\ ,\nonumber
\eea
where we introduced the shorthand $(\eta_1 ,...,\eta_n )(\xbo )=\big(\eta_1 (\xbo ),...,\eta_n (\xbo )\big)$. With this and a change of variables in the summation over $\xbo$ we obtain
\bea\label{dir1}
\langle\mathcal{L} f\rangle_{\bar{\nu}^L_\mubo } &=&\sum_{\feta\in\Omega_L}f(\feta )\sum_{i=1}^N \sum_{\xbo ,\ybo\in\Lambda_L}p_i (\ybo )\,\bar{\nu}_\mubo \big(\feta (\xbo\! -\!\ybo )\big)\,\bar{\nu}_\mubo \Big((\eta_1 ,..,\eta_i \! -\! 1,..,\eta_n )(\xbo )\Big)\nonumber\\
& &\qquad\left[ {g_i \Big((\eta_1 ,..,\eta_i \! +\! 1,..,\eta_n )(\xbo\! -\!\ybo )\Big)\,\bar{\nu}_\mubo \Big((\eta_1 ,..,\eta_i \! +\! 1,..,\eta_n )(\xbo\! -\!\ybo )\Big)\over\bar{\nu}_\mubo \big(\feta (\xbo\! -\!\ybo )\big) }-\right.\nonumber\\
& &\qquad\left.\ \ {g_i \big(\feta(\xbo )\big)\,\bar{\nu}_\mubo \big(\feta (\xbo )\big)\over\bar{\nu}_\mubo \Big((\eta_1 ,..,\eta_i \! -\! 1,..,\eta_n )(\xbo )\Big) }\right]\prod_{\zbo\in\Lambda_L \setminus\{\xbo -\ybo ,\xbo\}}\bar{\nu}_\mubo \big(\feta (\zbo )\big) =0\ ,
\eea
which vanishes by (\ref{cancel}) for every $f\in C(\Omega_L ,\rel )$. Thus (\ref{toshow}) is shown and $\bar{\nu}^L_\mubo$ is stationary.

Assume now that $\nu^L$ is an arbitrary stationary product measure of the zero range process with generator $\mathcal{L}$. Then
\bea\label{rel0}
\langle\mathcal{L} f\rangle_{\nu^L} =0\quad\mbox{for all}\quad f\in C(\Omega_L ,\rel )
\eea
and, by inserting special functions $f$, one deduces conditions on the jump rates. Consider a configuration $\bar{\feta}$ where there are $\kbo =(k_1 ,\ldots ,k_n )\in\nat^n$ particles at a fixed site $\xbo$ and the rest of the lattice is empty, i.e.\ $\bar{\feta}(\ybo )=\delta_{\ybo ,\xbo}\kbo$ for all $\ybo\in\Lambda_L$. From the stationarity condition (\ref{rel0}) with $f=\chi_{\bar{\feta}}$ we obtain for the one-point marginal of $\nu^L$
\bea\label{rel1}
\Big[ g_1 (\kbo ) +\ldots +g_n (\kbo )\Big] \,\nu (\kbo )=\sum_{i=1}^N \nu (\kbo^i ,k_i -1)\ ,
\eea
where we set $\nu (\kbo )=0$ if $k_i <0$ for some $i\in\{ 1,\ldots ,n\}$. To get (\ref{rel1}) we used
\bea\label{rel2}
\nu (0,..,0,k_i ,0,..,0 ) =\nu (\mathbf{0}) \prod_{l=1}^{k_i} {1\over g_i (0,..,0,l,0,..,0)}\ ,
\eea
as known from the stationary measure of the one component system (\ref{grandcanens}). Now let $f$ be the indicator function of $\bar{\feta}^{i;\xbo ,\xbo +\ybo}$, with $\bar{\feta}=\big(\delta_{\ybo ,\xbo}\kbo\big)_{\ybo\in\Lambda_L}$ as above and $k_i >0$. Using (\ref{rel0}) to (\ref{rel2}) we obtain
\bea\label{rel3}
\lefteqn{p_i (\ybo )\nu (\mathbf{0})\Big[ g_i (\kbo )\,\nu (\kbo )-\nu \big(\kbo^i ,k_i -1\big)\Big] =}\nonumber\\
& &\qquad =-\sum_{j\neq i} p_j (-\ybo )\,\nu \big(\kbo^{i,j} ,k_i -1,k_j -1\big)\Big[ g_j (\ebo_i +\ebo_j )\,\nu (\ebo_i +\ebo_j )\ -\nu (\ebo_j )\Big]\ .
\eea
with the shorthand $\kbo =(\kbo^{i,j} ,k_i ,k_j )$ and $\ebo_i \in\rel^n$ the unit vector in direction $i$. (\ref{rel3}) holds for all $\ybo\in\Lambda_L$ and is obviously fulfilled if the two square brackets vanish individually. Under the assumption that they do not vanish, one can easily construct a contradiction to (\ref{rel1}). Thus we obtain
\bea\label{rel4}
\nu (\kbo ) =\nu (\kbo^i ,k_i -1)/g_i (\kbo )
\eea
for all $i\in\{ 1,\ldots ,n\}$. Applying (\ref{rel4}) twice in different order for arbitrary $i\neq j$ we get
\bea\label{rel5}
\nu (\kbo )={\nu (\kbo^{i,j} ,k_i -1,k_j -1)\over g_i (\kbo )\, g_j (\kbo^i ,k_i -1)}={\nu (\kbo^{i,j} ,k_i -1,k_j -1)\over g_j (\kbo )\, g_i (\kbo^j ,k_j -1)}\ ,
\eea
and (\ref{a1}) easily follows.\hfill$\Box$\\
\\
\textbf{Remark:} If the jump rates are site dependent and satisfy (\ref{a2}) with potential $G_\xbo$ for every site $\xbo$, the measure defined analogous to (\ref{statmeas}) is still stationary, since the terms in (\ref{dir1}) cancel for each site individually. However, we did not see how to generalize the reverse argument to space-dependent rates.

  \subsection{Properties of the stationary measures}
The particle density $R_i$ of component $i\in\{ 1,\ldots ,n\}$ is translation invariant and given by
\bea\label{rho}
R_i (\mubo ):=\big\langle\eta_i (\xbo )\big\rangle_{\bar{\nu }_\mubo} =\big\langle\eta_i (\mathbf{0})\big\rangle_{\bar{\nu }_\mubo} =\partial_{\mu_i}\log Z(\mubo )\geq 0
\eea
as a function of the chemical potentials $\mubo$. For $\mubo\in\Di_\mu$ the measure $\bar{\nu}_\mubo$ has some finite exponential moments as $\langle e^{\boldsymbol{\theta}\cdot\feta (\mathbf{0})}\rangle_{\bar{\nu }_\mubo} =Z(\mubo +\boldsymbol{\theta})/Z(\mubo )<\infty$ for sufficiently small $\boldsymbol{\theta}\in\rel^n$. Therefore $\Rbo =(R_1 ,\ldots ,R_n ):\Di_\mu \to D_\rho$ is well defined with $D_\rho =\Rbo \big(\Di_\mu \big)\subset (0,\infty )^n$. Due to (\ref{rho}), $\log Z(\mubo )$ is monotonic increasing on $\Di_\mu$ and the compressibility is given by the matrix of second derivatives as
\bea
D\Rbo (\mubo )\! :=\!\Big(\partial_{\mu_j} R_i (\mubo )\Big)_{ij} \!\!\! =\! D^2 \log Z(\mubo )\! =\!\Big(\partial^2_{\mu_i \mu_j} \log Z(\mubo )\Big)_{ij} \!\!\! =\!\Big(\langle\eta_i (\mathbf{0})\eta_j (\mathbf{0})\rangle^c_{\bar{\nu }_\mubo}\Big)_{ij} ,
\eea
where $\big\langle\eta_i (\mathbf{0})\eta_j (\mathbf{0})\big\rangle^c_{\bar{\nu }_\mubo} :=\big\langle\eta_i (\mathbf{0})\eta_j (\mathbf{0})\big\rangle_{\bar{\nu }_\mubo} -\big\langle\eta_i (\mathbf{0})\big\rangle_{\bar{\nu }_\mubo} \big\langle\eta_j (\mathbf{0})\big\rangle_{\bar{\nu }_\mubo}$. Thus $D\Rbo (\mubo )$ is symmetric and positive definite, because
\bea
\abo^T \cdot\big( D^2 \log Z(\mubo )\big)\,\abo =\Big\langle\Big(\sum_{i=1}^n a_i \,\eta_i (\mathbf{0})\Big)^2\Big\rangle^c_{\bar{\nu }_\mubo} >0
\eea
for all $\abo\in\rel^n$ with $|\abo|=1$. Hence the eigenvalues of $D^2 \log Z(\mubo )$ are real and positive, which ensures that $\log Z(\mubo )$ is strictly convex and $\Rbo$ is invertible on $\Di_\mu$. Since $D\Rbo (\mubo )$ is also continuous, $D_\rho$ is diffeomorphic to $\Di_\mu$. We denote the inverse of $\Rbo$ by $\Mbo =(M_1 ,\ldots ,M_n ):D_\rho\to\Di_\mu$ and define the measure $\nu^L_{\rhobo } :=\bar{\nu}^L_{\Mbo (\rhobo )}$, which is indexed by the particle densities $\rhobo$. There exists an $\alpha >0$ such that $\rhobo\in D_\rho$ for all $|\rhobo |<\alpha$, so $\nu^L_{\rhobo }$ is defined for small densities. In many cases it is $D_\rho =(0,\infty )^n$, e.g.\ under the assumption (\ref{fem}) or (\ref{slg}) in Section 3.1. However, there are also cases where there is no stationary product measure for large densities. These systems show an interesting condensation phenomenon for large $\rhobo$ and have been studied for one-component systems in \cite{evans00,stefan}. In this case the behavior of $Z(\mubo )$ and $\Rbo (\mubo )$ at the boundary $\partial D_\mu$ is of importance. We will not discuss this point any further here.

The stationary current of component $i$ is given by
\bea\label{current}
\Jbo_i (\rhobo )=\mbo (p_i )\,\langle g_i \rangle_{\nu_{\rhobo }} =\mbo (p_i )\,\exp\big[ M_i (\rhobo )\big]
\eea
as a function of the particle densities. Here $\mbo (p_i )=\sum_{\ybo\in\Lambda_L} \ybo p_i (\ybo )\in\rel^d$ denotes the first moment of the jump probabilities $p_i$, which in general is non-zero and determines the direction of the current. The strength is proportional to $\exp\big[ M_i (\rhobo )\big]$ and thus a monotonic increasing function of the chemical potential.

The thermodynamic entropy $S(\rhobo )$ of the stationary measure is the convex conjugate of $\log Z(\mubo )$ given by the Legendre transform
\bea\label{entropy}
S(\rhobo )=\sup_{\mubo\in D_\mu}\bigg(\sum_{i=1}^n \rho_i \mu_i -\log Z(\mubo )\bigg)\ .
\eea
With (\ref{current}) we have for all $i\in\{ 1,\ldots ,n\}$
\bea\label{entrder}
\partial_{\rho_i} S(\rhobo )=M_i (\rhobo )=\log\langle g_i \rangle_{\nu_{\rhobo }}\ .
\eea
Therefore we have the following relation for the determinants, denoted by $|..|$,
\bea
\big| D^2 S(\rhobo )\big| =\big| D\Mbo (\rhobo )\big| =\big| D\Rbo (\Mbo (\rhobo ))\big|^{-1} =\big| D^2 \log Z(\Mbo (\rhobo ))\big|^{-1} >0\ .
\eea
Thus $S$ is strictly convex on $D_\rho$. Note that due to the structure of the stationary measure the densities are given as derivatives of the partition function with respect to the chemical potentials. This leads to
\bea
\partial_{\rho_j} \log\langle g_i \rangle_{\nu_\rhobo} =\partial_{\rho_i} \log\langle g_j \rangle_{\nu_\rhobo}\ ,
\eea
for all $i,j\in\{ 1,\ldots ,n\}$, which can be considered as the macroscopic analogue of condition (\ref{a1}) on the jump rates.

\section{Hydrodynamics}
  \subsection{The hydrodynamic limit}
We will show that under Eulerian scaling $t\to t/L$, $\xbo\to\ubo =\xbo /L$ in the limit $L\to\infty$ the time evolution of the local particle densities $\rhobo (t,\ubo )$ is given by the following system of conservation laws:
\bea\label{pde}
\partial_t \rho_i (t,\ubo) +\sum_{k=1}^d \partial_{u_k}J_i^k \big(\rhobo (t,\ubo )\big) =0\ ,\quad i=1,\ldots ,n\ ,
\eea
where $J_i^k \big(\rhobo\big)$ is the $k$-th spatial component of the $i$-th current $\Jbo_i (\rhobo )$ defined in (\ref{current}) and $\ubo =(u_1 ,\ldots ,u_d )\in \Lambda =\big(\rel/\ganz\big)^d$ is the continuous space variable. To prove the convergence, the dynamics of the zero range process has to be well defined in the limit $L\to\infty$ which is guaranteed by (see \cite{andjel82,kipnislandim})
\bea\label{extracond}
\sup_{i,j \in\{ 1,\ldots ,n\}}\sup_{k\in\nat} |g_i (\kbo^j ,k_j +1)-g_i (\kbo )|<\infty\ .
\eea
We also need to impose an extra condition on the stationary measure, which is needed for the one block estimate (see \cite{kipnislandim}, Chapter 5), as one important part of the convergence proof. There are two alternatives, the first one is to require that the partition function $Z(\mubo )$ is finite for all $\mu\in\rel^n$, which is equivalent to the existence of finite exponential moments, i.e.
\bea\label{fem}
\big\langle\exp [\boldsymbol{\theta }\cdot\feta (\mathbf{0})] \big\rangle_{\bar{\nu}_\mubo} =Z(\mubo +\boldsymbol{\theta})/Z(\mubo )<\infty\mbox{ for all }\boldsymbol{\theta}\in\rel^n \quad\Leftrightarrow\quad D_\mu =\rel^n \ .
\eea
Note that this implies $D_\rho =(0,\infty )^n$, avoiding possible problems in case $\rhobo (t,\ubo )$ reaches the boundary of $D_\rho$. Alternatively, we can impose sublinearity of the jump rates, i.e.\ for all $i\in\{ 1,\ldots ,n\}$ and all $\mathbf{b}\in (0,\infty )^n$ there exist $a_i (\mathbf{b} )\in\rel$ such that
\bea\label{slg}
g_i (\kbo )\leq a_i (\mathbf{b} ) +\mathbf{b}\cdot\kbo\quad\mbox{and}\quad\lim_{\mubo\to\mubo^*} Z(\mubo )=\infty \mbox{ for all }\mubo^* \in\partial D_\mu \ .
\eea
The second statement is needed to ensure $D_\rho =(0,\infty )^n$, since sublinearity does not rule out $D_\mu \subsetneqq\rel^n$. Given a solution $\rhobo (t,\ubo )$ of (\ref{pde}) we denote the corresponding local equilibrium measure by
$\nu^L_{\rhobo (t,.)}$, which will be compared to the time dependent distribution $\pi_t^L$ of the zero range process,
\bea\label{makro}
\nu^L_{\rhobo (t,.)} :=\prod_{\xbo\in\Lambda_L} \nu_{\rhobo (t,\xbo /L)}\qquad\mbox{and}\qquad\pi_t^L =\pi_0^L S_{tL}
\eea
where $S_{tL}$ is the semi-group $S_t$ associated to the generator $\mathcal{L}$ speeded up by $L$. The proof of the following theorem is an application of Yau's relative entropy method \cite{yau91}, which requires some regularity of the solution $\rhobo (t,\ubo )$. In general solutions of conservation laws develop shocks after a finite time even for smooth initial data (see e.g.\ \cite{smoller}). Thus the convergence proof is valid only up to the time $T$ of the appearance of the first discontinuity.\\
\\
\textbf{Theorem 2:} (Hydrodynamic limit)\it\\
Let $\rhobo\in C^2 \big( [0,T]\times\Lambda ,[0,\infty )^n \big)$ be a solution of (\ref{pde}) for some $T\in (0,\infty )$ with smooth and bounded initial profile $\rhobo (0,.)$, satisfying $\rho_i (0,\ubo )\geq\rho_i^* >0$, $i=1,\ldots ,n$. Under the assumptions (\ref{extracond}) and (\ref{fem}) resp.\ (\ref{slg}) let $\pi_0^L$ be a sequence of probability measures on $\Omega_L$, whose entropy $H\big(\pi_0^L \big|\nu^L_{\rhobo (0,.)} \big)$ relative to $\nu^L_{\rhobo (0,.)}$ is of order $o(L^d )$. Then
\bea\label{result2}
H\big(\pi_t^L \big|\nu^L_{\rhobo (t,.)} \big) =o(L^d )\quad\mbox{for all }\ t\in [0,T]\ .
\eea\rm\\[-5mm]

Applying the entropy inequality in the standard way (see \cite{kipnislandim}, Chapter 6), Theorem 2 implies the following\\
\\
\textbf{Corollary:}\it\ Under the assumption of Theorem 2, for any smooth test function $f:\Lambda \to\rel$, $t\in [0,T]$ and $i=1,\ldots ,n$, the following limit
\bea
\lim_{L\to\infty} {1\over L^d} \sum_{\xbo\in\Lambda_L} f(\xbo /L)\,\eta_i (t,\xbo )=\int_{\Lambda} f(\ubo )\,\rho_i (t,\ubo )\, d^d u
\eea
holds in probability, where $\feta (t,\xbo )$ denotes the time $t$ configuration of the zero range process with distribution $\pi_t^L$.
\rm\\

The proof of Theorem 2 is close to the ones given in \cite{kipnislandim,tothetal02} and its most important steps will be sketched in Section 3.3.

  \subsection{Properties of the limit equation}
Introducing the matrices $D\Jbo^k (\rhobo ):=\big(\partial_{\rho_j} J_i^k \big)_{ij}$ one can rewrite (\ref{pde}) in the quasilinear form
\bea\label{quasilinear}
\partial_t \rhobo (t,\ubo)+\sum_{k=1}^d D\Jbo^k \big(\rhobo (t,\ubo )\big)\,\partial_{u_k} \rhobo (t,\ubo )=0\ .
\eea
The current is given by
\bea\label{deltadef}
D\Jbo^k (\rhobo )=\Delta^k (\rhobo )\, D\Mbo (\rhobo )\quad\mbox{with}\quad\Delta^k_{ij} (\rhobo )=\delta_{ij}\, m_k (p_i )\,\exp\big[ M_i (\rhobo )\big]\ ,
\eea
where $\Delta^k$ is a diagonal matrix and $m_k (p_i )$ is the $k$-th space component of $\mbo (p_i )$. Since $D\Mbo (\rhobo )=D^2 S(\rhobo )$ is symmetric and positive definite, we can write for all $k=1,\ldots ,d$
\bea\label{symsim}
D\Jbo^k =\big(D^2 S \big)^{-1/2} \Big(\big(D^2 S \big)^{1/2} \Delta^k \big(D^2 S \big)^{1/2}\Big)\big(D^2 S \big)^{1/2}\ .
\eea
Thus $D\Jbo^k$ is similar to the real symmetric matrix $\big(D^2 S \big)^{1/2} \Delta^k \big(D^2 S \big)^{1/2}$, which implies that $\sum_{k=1}^d \omega_k D\Jbo^k$ is diagonalizable for all $\boldsymbol{\omega}=(\omega_1 ,\ldots ,\omega_d )\in\rel^d$ with $|\boldsymbol{\omega}|=1$ and that (\ref{pde}) is hyperbolic \cite{ulbrich}. The question of strict hyperbolicity, i.e.\ whether all eigenvalues of $\sum_{k=1}^d \omega_k D\Jbo^k$ are nondegenerate, cannot be answered in general. It depends on the dynamics of the zero range process and one has to check for each system separately.

With (\ref{entrder}) and (\ref{deltadef}) it is easy to see that $S(\rhobo )$ satisfies
\bea\label{eefpair}
\sum_{i=1}^n \partial_{\rho_i} S(\rhobo )\, D\Jbo_{ij}^k (\rhobo )=\partial_{\rho_j} F_k \big(\Mbo (\rhobo )\big)\quad\mbox{for all }j=1,\ldots ,n,\ k=1,\ldots ,d\ ,
\eea
provided we set $\Fbo (\mubo )=\sum_{j=1}^n \mbo (p_j )\,\mu_j \,\exp [\mu_j ]$. (\ref{eefpair}) are the $dn$ defining relations for entropy entropy-flux pairs of hyperbolic systems \cite{smoller,ulbrich}. For general systems with $dn>d+1$ these equations are overdetermined and the existence of an entropy entropy-flux pair is not guaranteed. However, as we just have shown, for a zero range process the Euler equation has always a strictly convex entropy $S$, defined in (\ref{entropy}), with corresponding flux $\Fbo$.

Systems of conservation laws with entropy are studied in detail in \cite{ulbrich}. In general, by transformation to the so-called entropy variables the quasilinear equation (\ref{quasilinear}) simplifies to a symmetric system. In our case these variables are given by the chemical potentials $\mubo (t,\ubo ):=\Mbo \big(\rhobo (t,\ubo )\big)$ and the derivative of the current with respect to $\mubo$ is even diagonal,
\bea\label{epde}
D\Rbo \big(\mubo (t,\ubo )\big)\,\partial_t \mubo (t,\ubo )+\sum_{k=1}^d \Delta^k \big(\mubo (t,\ubo )\big)\,\partial_{u_k} \mubo (t,\ubo ) =0\ ,
\eea
where $\Delta^k (\mubo ):=\Delta^k \big(\Rbo (\mubo )\big)$ is defined in (\ref{deltadef}).

To summarize we have shown that the limit equation (\ref{pde}) is a hyperbolic system with globally convex entropy. General results on the existence and uniqueness of solutions of such systems are rare, even for the simple form (\ref{epde}). So further results for systems with open boundaries are based on solid arguments rather than rigorous proofs and presented in Section 4.

  \subsection{Proof of Theorem 2}
The proof follows closely the one given in \cite{kipnislandim}, Chapter 6, and \cite{tothetal02} and we only sketch the main steps. The only part of the proof where the structure of the stationary measure for $n$-component systems enters is (\ref{block2}), where we use the symmetry of $D\Mbo (\rhobo )$. Since $\pi_t^L$ and $\nu^L_{\rhobo (t,.)}$ are absolutely continuous with respect to each other and with respect to a reference invariant measure $\nu^L_\falpha$, $\falpha\in (0,\infty )^n$, one can define the density
\bea
\psi^L_t (\feta ):={d \nu^L_{\rhobo (t,.)}\over d\nu^L_\falpha}=\prod_{\xbo\in\Lambda_L} {Z\big(\Mbo (\falpha )\big)\over Z\big(\Mbo (\rhobo (t,\xbo /L))\big)}\prod_{i=1}^n {\exp \big[\eta_i (\xbo ) M_i \big(\rhobo (t,\xbo /L)\big] \over\exp \big[\eta_i (\xbo ) M_i (\falpha )\big] }\ .
\eea
Let $H_L (t):=H\big(\pi_t^L \big|\nu^L_{\rhobo (t,.)}\big)$ be the entropy of $\pi_t^L$ relative to $\nu^L_{\rhobo (t,.)}$. To establish the estimate (\ref{result2}), we will prove a Gronwall type inequality
\bea\label{gronwall}
H_L (t)\leq H_L (0)+C\int_0^t H_L (s)\, ds+o(L^d )
\eea
with uniform error bound for all $t\in [0,T]$. The entropy production is bounded above by
\bea\label{hprod}
\partial_t H_L (t)\leq\int_{\Omega_L} {1\over \psi_t^L (\feta )}\Big( L\mathcal{L}^* \psi_t^L (\feta )-\partial_t \psi_t^L (\feta )\Big)\, d\pi_t^L\ ,
\eea
where $\mathcal{L}^*$ is the adjoint of $\mathcal{L}$ in $L^2 (\nu_\falpha^L )$. This inequality is proved in \cite{kipnislandim}, Chapter 6, under very general conditions covering our case. Using the regularity of $\Mbo \big(\rhobo (t,.)\big)$, the right hand side of (\ref{hprod}) can be rewritten as
\bea\label{zwerg}
\lefteqn{\big(\psi_t^L (\feta)\big)^{-1} L\mathcal{L}^* \psi_t^L (\feta )=-\sum_{\xbo\in\Lambda_L}\sum_{i=1}^n \sum_{k=1}^d \partial_{u_k} J_i^k \big(\rhobo (t,\xbo /L) \big) }\nonumber\\
& &\qquad -\sum_{\xbo\in\Lambda_L}\sum_{i=1}^n \sum_{k=1}^d \partial_{u_k} M_i \big(\rhobo (t,\xbo /L) \big)\Big( m_k (p_i )\, g_i \big(\feta (\xbo )\big) -J_i^k \big(\rhobo (t,\xbo /L) \big) \Big) +\ord (L^{d-1})\nonumber\\
\lefteqn{\big(\psi_t^L (\feta)\big)^{-1} \partial_t \psi_t^L (\feta)=\sum_{\xbo\in\Lambda_L}\sum_{i=1}^n \partial_t \, M_i \big(\rhobo (t,\xbo /L) \big)\Big(\eta_i (\xbo )-\rho_i (t,\xbo /L) \Big)\ .}
\eea
The right hand side of the first line is a telescoping sum and vanishes up to an error $\ord (L^{d-1})$. Because of the regularity of $\Mbo (\rhobo )$ a summation by parts permits to replace the local variables by their block averages. They are defined as
\bea\label{block}
\eta_i^\ell (\xbo )={1\over (2\ell +1)^d}\sum_{|\xbo -\ybo|\leq\ell } \eta_i (\ybo )\ ,\quad g_i^\ell \big(\feta (\xbo)\big) ={1\over (2\ell +1)^d}\sum_{|\xbo -\ybo|\leq\ell } g_i \big(\feta (\ybo)\big) 
\eea
for $i=\{1,\ldots ,n\}$ and $\ell\in\mathbb{Z}^+$. Using the hyperbolic system (\ref{pde}) and the symmetry of $D\Mbo$ we obtain
\bea\label{block2}
\partial_t M_i \big(\rhobo (t,\xbo /L)\big) =-\sum_{j=1}^n \sum_{k=1}^d \partial_{\rho_i} J_j^k \big(\rhobo (t,\xbo /L)\big)\,\partial_{u_k} M_j \big(\rhobo (t,\xbo /L)\big)\ .
\eea
Inserting (\ref{block}) and (\ref{block2}) in (\ref{zwerg}),
\bea\label{zwerg2}
\lefteqn{\int_{\Omega_L} {1\over \psi_t^L}\Big(\partial_t \psi_t^L \! -\! L\mathcal{L}^* \psi_t^L \Big)\, d\pi_t^L =\int_{\Omega_L} \sum_{\xbo\in\Lambda_L}\sum_{i=1}^n \sum_{k=1}^d \Big( m_k (p_i )\, g_i^\ell \big(\feta (\xbo )\big)\! -\! J_i^k \big(\feta^\ell (\xbo )\big)\!\Big)\, d\pi_t^L}\nonumber\\
& &\qquad\qquad +\int_{\Omega_L} \sum_{\xbo\in\Lambda_L}\sum_{i=1}^n \sum_{k=1}^d \partial_{u_k} M_i \big(\rhobo (t,\xbo /L) \big)\ f_i^k \Big(\feta^\ell (\xbo ),\rhobo (t,\xbo /L)\Big)\, d\pi_t^L \ ,
\eea
where
\bea
f_i^k \big(\abo ,\bbo\big) =J_i^k (\abo )-J_i^k (\bbo )-\grad J_i^k (\bbo )\cdot (\abo -\bbo )\ .
\eea
A bound for the second term on the right hand side of (\ref{zwerg2}) comes from the entropy inequality (see \cite{kipnislandim}, Chapter 6). Dropping the argument of $f_i^k$ we get
\bea\label{bound1}
\int_{\Omega_L} \sum_{\xbo\in\Lambda_L}\sum_{i=1}^n \sum_{k=1}^d \partial_{u_k} M_i \big(\rhobo (t,\xbo /L) \big)\,\big| f_i^k \big|\, d\pi_t^L \leq C H\big(\pi_t^L \big|\nu^L_{\rhobo (t,.)}\big) +\ord (L^d \ell^{-1})\ .
\eea
The first term is estimated integrated in time by using the so-called one block estimate (see \cite{kipnislandim}, Chapter 5)
\bea\label{bound2}
\lim_{\ell\to\infty} \lim_{L\to\infty} L^{-d} \int_0^t \int_{\Omega_L} \sum_{\xbo\in\Lambda_L}\sum_{i=1}^n \sum_{k=1}^d \Big( m_k (p_i )\, g_i^\ell \big(\feta (\xbo )\big)\! -\! J_i^k \big(\feta^\ell (\xbo )\big)\!\Big)\, d\pi_s^L\, ds=0\ .
\eea
This is the only part of the proof where either one of the two regularity assumptions on the stationary measure (\ref{fem}) or (\ref{slg}) is used. Inserting (\ref{bound1}) and (\ref{bound2}) in (\ref{zwerg2}) we obtain (\ref{gronwall}) via (\ref{hprod}) and Theorem 2 follows.

\section{Stationary solutions for systems with open boundaries}

  \subsection{Uniqueness criterion for the physical solution}
In this section we turn to one of the motivations of our study and apply the results of the previous sections to determine stationary density profiles of systems with open boundary conditions. General results on existence and uniqueness of solutions to systems of hyperbolic conservation laws like (\ref{pde}) and (\ref{epde}) are not available and we have no other choice than to base our results on solid arguments rather than rigorous proofs. We explicitly state our assumptions.

Although (\ref{epde}) is derived only up to the first discontinuity, we assume in the following the validity of the Euler equation in the sense of weak solutions. They are in general not unique and we have to find a criterion to single out the physical solution, i.e.\ the one which describes density profiles of the underlying zero range process. There has been a lot of work on this problem (see e.g.\ \cite{smoller}) and one possibility is to add a viscosity term with a small parameter $\epsilon$ on the right hand side of (\ref{epde}). A natural choice of this term is the diffusive correction which was neglected in the derivation of (\ref{pde}), where the small parameter is interpreted as $\epsilon =\ord (L^{-d} )$. In our context this term is given by the Green-Kubo formula for the corresponding reversible zero range process with symmetric jump probabilities (see \cite{spohn}, Chapter II.2.2). Since the symmetric zero range process is a gradient system, the viscosity is already determined by the stationary measure itself. We get the following dissipative equation
\bea\label{viscous}
\sum_{j=1}^n DR_{ij} (\mubo^\epsilon )\,\partial_t \mu_j^\epsilon +\sum_{k=1}^d m_k (p_i )\,\partial_{u_k} \exp [\mu_i^\epsilon ] =\epsilon\sum_{k,l=1}^d \sigma_{kl} (p_i )\,\partial^2_{u_k u_l} \exp [\mu_i^\epsilon ]\ ,
\eea
for $i=1,\ldots ,n$, with $\sigma_{kl} (p_i )=\sum_{\xbo\in\Lambda_L} x_k x_l \, p_i (\xbo )$. The solution $\mubo^\epsilon (t,\ubo )$ of Equation (\ref{viscous}) is unique and, if it converges in a proper sense for $\epsilon\searrow 0$, the limit $\mubo^0 (t,\ubo )$ is a weak solution of (\ref{epde}). This convergence is in general difficult to show and in the following we will just assume that it holds, for details see \cite{ulbrich} Chapter 3.8 and references therein.

Since $D^2 S (\rhobo )\Big(\delta_{ij}\sigma_{kl}(p_i) \partial_{\rho_j}\exp\big[ M_i (\rhobo )\big]\Big)_{ij}$ is similar to a symmetric, positive definite matrix (analogous to (\ref{symsim})), we have for some $\delta >0$ and arbitrary $\abo_k \in\rel^n$, $k=1,\ldots ,d$,
\bea
\sum_{k,l=1}^d \abo_k^T D^2 S (\rhobo )\Big(\delta_{ij}\sigma_{kl}(p_i)\,\partial_{\rho_j}\exp\big[ M_i (\rhobo )\big]\Big)_{ij} \abo_l \geq\delta \sum_{k=1}^d \|\abo_k \|_2^2 \geq 0\ .
\eea
This expresses the viscous dissipation for the entropy and ensures (see \cite{ulbrich}, Chapter 3.8) that $\mubo^0 (t,\ubo )$ satisfies the entropy inequality
\bea
\partial_t S\big(\Rbo (\mubo^0 (t,\ubo ))\big) +\sum_{k=1}^d \partial_{u_k} F_k \big(\mubo^0 (t,\ubo )\big) \leq 0\ ,
\eea
as to be expected for a physical solution. Thus we expect that the zero viscosity limit of (\ref{viscous}) describes the macroscopic chemical potential profiles of the zero range process. Note that for stationary solutions the system (\ref{viscous}) decouples and stationary profiles can be obtained very easily. They are only determined by the first and second moment of the jump probabilities and the boundary conditions, whereas they are independent of the jump rates. The rates only enter the partition function $Z(\mubo )$ and thus the transformation to density profiles via $\Rbo (\mubo )$ (\ref{rho}), which is illustrated in the next section.

  \subsection{Stationary profiles for one-dimensional systems}
There has been considerable activity to understand the structure of the nonequilibrium steady state of systems with open boundaries. Here we study this issue in one dimension on the level of the hydrodynamic equations. It turns out that the fugacity variables $\phi_i :=\exp [\mu_i ]$ are the most convenient choice. As a consequence of (\ref{viscous}), for $d=1$ the stationary fugacity profiles $\phi^0_i (u)$, $u\in [0,1]$ are the limit solutions for $\epsilon\searrow 0$ of the equation
\bea\label{1dequ}
m (p_i )\,\partial_u \phi_i^\epsilon (u) =\epsilon\,\sigma (p_i )\,\partial^2_u \phi_i^\epsilon (u)\ ,\quad i=1,\ldots ,n\ .
\eea
The equations are decoupled and for a system with open boundary conditions $\phi_i (u)=\exp \big[ M_i (\rhobo (u))\big]$ for $u=0,1$, the solution is given by
\bea\label{phisol}
\phi^\epsilon_i (u) =\phi_i (0)+\big(\phi_i (1)-\phi_i (0)\big)\big( q_i (\epsilon )^u -1\big)\big/\big( q_i (\epsilon ) -1\big)\ ,
\eea
where $q_i (\epsilon )=\exp \big[ m(p_i )/(\epsilon\sigma (p_i ))\big]$. The profiles (\ref{phisol}) have a very simple structure. They are bounded above and below by $\phi_i (0)$ resp.\ $\phi_i (1)$ and for $\epsilon\searrow 0$ they converge pointwise to flat curves $\phi^0_i (u)$ with a jump at one of the boundaries, if $\phi_i (0)\neq \phi_i (1)$. The location of the jump depends on the sign of $m(p_i )$, which corresponds to the direction of the current. The coupled transformation to the stationary density profile $\rho_i^* (u)$ involves the fugacities of all components and is given by $\rho_i^0 =R_i \big(\log\phi^0_1 ,\ldots ,\log\phi^0_n \big)$ defined in (\ref{rho}).

\bef
\begin{center}
\includegraphics[width=\textwidth]{profiles.epsi}
\end{center}
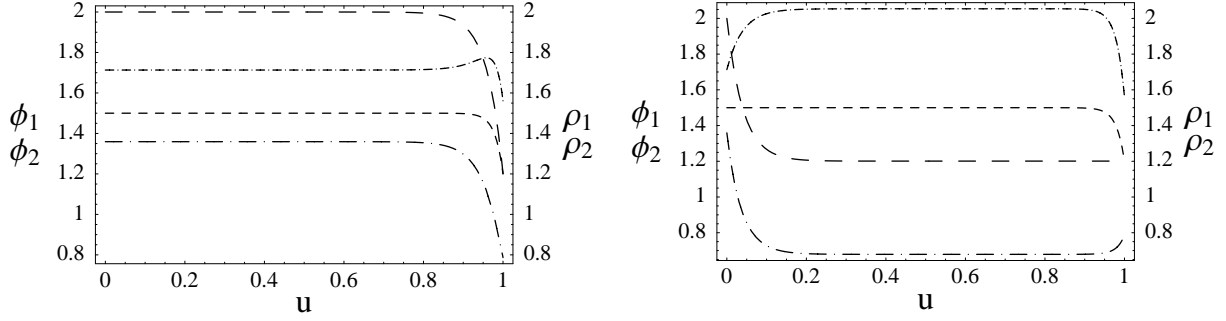
\caption{\small Stationary solution of (\ref{1dequ}) with $\epsilon =0.02$ for a one-dimensional two-component system with open boundary conditions $\phi_1 (0)=1.5$, $\phi_1 (1)=1.2$, $\phi_2 (0)=2$, $\phi_2 (1)=1.2$ and jump rates (\ref{exrates}) with $c_1 =1.3$, $c_2 =1$. Fugacities: dashed lines $\phi_1$ (- - -), $\phi_2$ (-- -- --) given by (\ref{phisol}), densities: dash-dotted lines $\rho_1$ (-$\cdot$-$\cdot$-), $\rho_2$ (--$\cdot$--$\cdot$--). Left: $m(p_1 )=1$, $m(p_2 )=0.5$. Right: $m(p_1 )=1$, $m(p_2 )=-0.5$.}
\label{profiles}
\enf
We illustrate our result with a simple example with two components, for which the jump rates are given by the potential
\bea\label{exrates}
G(k_1 ,k_2 )=k_1\log c_1 +k_2 \log c_2 +\log (k_1 +k_2 )!
\eea
and thus read $g_i (k_1 ,k_2 )=\chi_{k_i >0}\, c_i \, (k_1 +k_2 )$, $i=1,2$. The two types of particles just move with different speeds, but in this case one can calculate the partition function (\ref{part}) analytically. In terms of the fugacities $\phi_1$ and $\phi_2$ it is given by
\bea
Z(\phi_1 ,\phi_2 )=\left\{\begin{array}{ccl}{\phi_1 /c_1 \exp[\phi_1 /c_1 ]-\phi_2 /c_2 \exp[\phi_2 /c_2 ]\over\phi_1 /c_1 -\phi_2 /c_2 }&,&\mbox{ for }\phi_1 /c_1 \neq \phi_2 /c_2 \\ \exp[\phi_1 /c_1 ](1+\phi_1 /c_1 )&,&\mbox{ for }\phi_1 /c_1 = \phi_2 /c_2 \ea\right.\ .
\eea
The transformation to densities is then obtained via $R_i (\phi_1 ,\phi_2 )=\phi_i \partial_\phi Z(\phi_1 ,\phi_2 )$ (cf.\ (\ref{rho})). The jump probabilities are chosen such that $\sigma (p_1 )=\sigma (p_2 )=1$ and $m(p_1 )=1$. Therefore the bulk value $\phi_1^0 (u)$, $u\in (0,1)$ is equal to the left boundary $\phi_1 (0)$. For $m(p_2 )=0.5$ the same is true for $\phi_2$ and also the density profiles are determined by their left boundary value. This can be seen in Figure 1 (left), where we plot the profiles for small $\epsilon =0.02$, for better illustration. For $m(p_2 )=-0.5$ the particle species are driven in opposite directions, leading to a combination of fugacities in the bulk, which at neither of the two boundaries is present. So the bulk densities no longer agree with their boundary values, cf.\ Figure 1 (right).

  \subsection{Concluding remarks to steady-state selection}
The eigenvalues $\lambda_i (\rhobo )$ of the current derivative $D\Jbo (\rhobo )$ defined in (\ref{deltadef}) determine the characteristic velocities at which small perturbations of a flat fugacity profile propagate. For zero range processes the sign of $\lambda_i (\rhobo )$ is fixed by the sign of the first moment of the jump probabilities $m(p_i )$ and independent of the densities $\rhobo$. So no matter how the boundary conditions are chosen, the qualitative behavior of the stationary profiles does not change. This is in contrast to one-dimensional driven lattice gases with exclusion dynamics, where a change of sign of the characteristic velocities is the key ingredient for boundary induced phase transitions (see \cite{krug91,popkovetal99}). So our analysis shows that these phenomena are not present in zero range processes and the selection of stationary states in terms of fugacities is particularly simple. The construction of stationary profiles as shown above can be readily generalized to higher space dimensions. Despite the absence of boundary phase transitions, the zero range process is a very important interacting particle system, last but not least due to its close relation to exclusion models \cite{evans00}. It is one of the few examples of multi-species systems, where the selection of stationary states is well understood under very general conditions.

\section*{Acknowledgments}
The authors would like to thank B\'alint T\'oth and Christian Klingenberg for valuable advice on PDEs. S.G.\ acknowledges the support of the Graduiertenkolleg ``Mathematik im Bereich ihrer Wechselwirkung mit der Physik'', of the Institut Henri Poincar\'e, and of the DAAD/CAPES program ``PROBRAL''.


\end{document}